# INDONESIAN'S PRESIDENTIAL SOCIAL MEDIA CAMPAIGNS

**Leon Andretti Abdillah**[1)]
[1]Information Systems Study Program, Computer Science Faculty, Bina Darma University
Jl. Ahmad Yani No.12, Palembang, 30264
Phone : (+62-711) 515679, Fax : (+62-711) 515581
E-mail : leon.abdillah@yahoo.com[1)]

### Abstract

*Social media has been used for political campaigns of presidential candidates in a number of democratic countries, including Indonesia.The candidates for president and vice president have been following the latest trends in the virtual world by using social media such as: 1) Facebook, and 2) Twitter.The author's continuing research with a focus on the popularity of candidates for president and vice president of the Indonesian presidential election in July 9, 2014.The authors found that Facebook remains a social media most widely used in Indonesia for presidential campaigns.Prabowo secured the most number of fans on Facebook. While Joko Widodo secured the most number of followers on Twitter.Facebook users is dominated by the user in the age range 18-24 years, and most cities are using Facebook is Jakarta.*

*Kata kuncis:Social media, Facebook, Twitter, President candidates, Presidential campaigns.*

### Abstrak

*Media sosial telah digunakan untuk kampanye politik calon presiden di sejumlah negara-negara demokrasi, tidak terkecuali Indonesia. Para calon presiden dan wakil presiden telah mengikuti tren terkini di dunia maya dengan memanfaatkan media sosial seperti: 1) Facebook, dan 2) Twitter. Penulis meneruskan penelitian ini dengan fokus popularitas calon residen dan wakil presiden menuju pemilihan presiden Indonesia pada 9 Juli 2014. Penulis menemukan bahwa Facebook tetaplah menjadi media sosial yang paling banyak digunakan di Indonesia untuk kampanye presiden. Prabowo Subianto berhasil mendapatkan jumlah penggemar terbanyak di Facebook. Sedangkan Joko Widodo berhasil mendapatkan jumlah pengikut terbanyak di Twitter.Pengguna facebook didominasi oleh pengguna pada rentang usia 18-24 tahun, dan kota terbanyak yang menggunakan Facebook adalah Jakarta.*

*Kata kuncis: Media sosial, Facebook, Twitter, Calon presiden, Kampanye presiden.*

## 1. INTRODUCTION

Among various of popular internet applications, social media is one of them [1]. These internet sites attract billions of cyber users, specially young adults. The rapid development of online social networks has tremendously changed the way of people to communicate with each other [2]. They change how people promoting their products [3], learning environment, personal communication, and knowledge dissemination [4]. And now, social networking (SN) sites are already conventional communications venues for young adults [5].

This article is a continuation of the discussion of the two previous articles [1][6]. Focus topic in this article covers presidential candidates campaigns by using social media. Some articles have discussed the presidential campaign in social media, among others: 1) Leadership, party, and religion: Explaining voting behavior in Indonesia [7], 2) Facebook and vote share in the 2008 presidential primaries [8], 3) Political engagement of young adults[5], 4) Predicting Elections with Twitter [9], and 5) Facebook,Twitter and YouTube Democratizing Our Political Systems [10].

Indonesia is one of the important emerging powers in Asia-Pacific, the world's third-largest democracy [11] after India and the United States, the world's largest majority-Muslim country[12], a member of the G-20,a country with one-third of ASEAN's GDP and almost 40 percent of its population[13]. After successful with the legislative elections held in 9 April 2014, Indonesia will next elect for the seventh president on July 9, 2014. In the 2014 presidential election, there are only two candidates who will compete, namely : 1) Prabowo Subianto + Hatta Rajasa, and 2) Joko Widodo + Jusuf Kalla. Both candidates are well known by the Indonesian. Both candidates have been used social media for their political campaigns.

Per June 2014, there are five top countries on facebook, and Indonesia in the position number four [14]. The top country of facebook users is USA, followed by India and Brazil. After Indonesia there is Mexico. Facebook users in Indonesia are dominated by young adult people (18-24), then adults people (25-34), see figure 1.

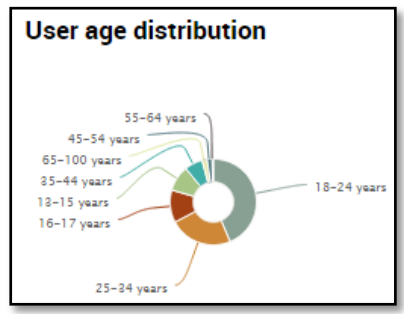

*Figure 1. Facebook User Age Distribution (SocialBakers, June 2014)*

Among all Facebook users in Indonesia, more male users than female users (figure 2). This condition is also the case of India, Brazil, and Mexico except the USA. In Indonesia, there are 41% female users compare to 50% of male users.

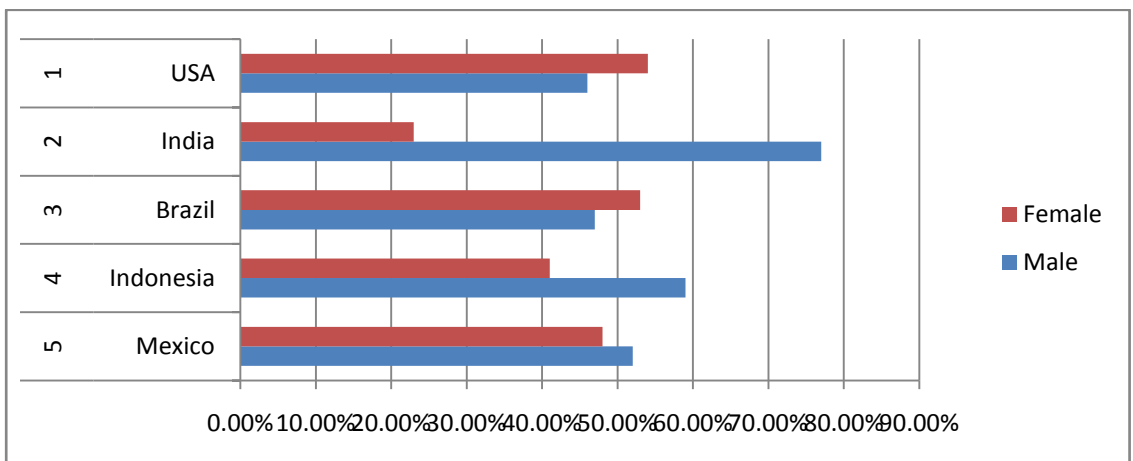

*Figure 2. Facebook User Gender Distribuiion (SocialBakers, June 2014)*

The article is divided into four sections: The first section situates the article within the dominant themes of the literature. The second is methodological. The next sections discuss the Indonesian presidential candidates social media. Finally, the article concludes with some points related to Indonesian's presidential political campaigns particular use of social media in the 2014 general election.

## 2. RESEARCH METHODS

In this article author add one social media, YouTube, for the observation. Then in this article author observes the features of political parties' social media, such as: 1) Facebook, 2) Twitter, and 3) YouTube. Author explores political parties' social media to check their activities. Author also gathers valuable information from KPU for the real count results (2009 and 2014).

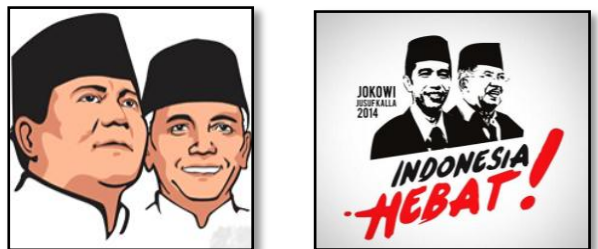

*Figure 3. Indonesian Presidential Candidates 2014*

### 2.1 Indonesian Presidential Candidates 2014

In presidential election 2014 there are only two sets of candidates [15] for president and vice president candidates: 1) Prabowo Subianto – Hatta Rajasa, and 2) Joko Widodo – Jusuf Kalla. Figure 3 shows the icon of both president candidates.

Table 1 shows the presidential candidates with political parties supporters. Candidates number 1 was supported by seven political parties: 1) Gerindra, 2) PAN, 3) Golkar, 4) PKS, 5) PPP, 6)PBB, dan PDemokrat. On the other hand, candidates number 2 was supported by five political parties: 1) PDIP, 2) Nasdem, 3) PKB, 4) PHanura, and 5) PKPI.

Table 1. Pair of Candidates for President and Vice President in the 2014 General Election

| No | Candidates | Political Parties Supporters | Percentation | Total |
|---|---|---|---|---|
| 1 | Prabowo Subianto – Hatta Rajasa | Gerindra | 11.81% | 59.12% |
|  |  | PAN | 7.59% |  |
|  |  | Golkar | 14.75% |  |
|  |  | PKS | 6.79% |  |
|  |  | PPP | 6.53% |  |
|  |  | PBB | 1.46% |  |
|  |  | Demokrat | 10.19% |  |
| 2 | Joko Widodo – Jusuf Kalla | PDIP | 18.95% | 40.88% |
|  |  | Nasdem | 6.72% |  |
|  |  | PKB | 9.04% |  |
|  |  | PHanura | 5.26 % |  |
|  |  | PKPI | 0.91% |  |

## 2.2 President Candidates' Social Media Accounts

Author visited social media used by the candidates: 1) Facebook, and 2) Twitter.Table 3 shows the presidential candidates'social media account. All of social media accounts use the full name or initial name of every president candidates.

Table 2. Social Media Accounts of Pair of Candidates for President and Vice President in the 2014 General Election

| No | Candidates | FaceBook | Twitter |
|---|---|---|---|
| 1 | Prabowo Subianto | PrabowoSubianto | @Prabowo08 |
| 2 | Joko Widodo | JKWofficial | @jokowi_do2 |
| 3 | Hatta Rajasa | M.Hatta.Rajasa | @hattarajasa |
| 4 | Jusuf Kalla | MUH-JUSUF-KALLA | @Pak_JK |

## 2.3 Statistics from Online Research Media

Authorincludes the research from SocialBakers. This is an important online statistics tool for researchers who want to read the statistics condition from every country. SocialBakers works with five major social media at the moment: 1) Facebook, 2) Twitter, 3) YouTube, 4) Google+, and 5) LinkedIn. But for this research, author only uses Facebook and Twitter.

SocialBakers is a user friendly social media analytics platform which provides a leading global solution that allows brands to measure, compare, and contrast the success of their social media campaigns with competitive intelligence [16]. SocialBakers also provide eleven clasifications, such as: 1) By Country, 2) Pages, 3) Brands, 4) Media, 5) Entertainment, 6) Sport, 7) Celebrities, 8) Society, 9) Community, 10) Places, and 11) Apps & Developers.

## 3. RESULTS AND DISCUSSIONS

Section three informs the results from 1) KPU's real count, 2) president candidates's facebook page, 3) president candidates's twitter account, and 4) social media as president candidates's virtual society in presidential campaigns 2014. The discussions are based on information before the presidential election.

### 3.1 Real Count Results for Political Party's Campaigns

KPU released the results of Indonesia general election 2014 on June 2014. Four big winners (>10%) are : 1) PDIP, 2) PGolkar, 3) PGerindra, and 4) PDemokrat. And for the president election 2014, the candidates are from : 1) PDIP (Jokowi), PGerindra (Prabowo), and PAN (Hatta Rajasa). Jusuf Kalla is PGolkar party cadres, but in the 2014 presidential election is not full supported by PGolkar.

Table 4. The List of Political Parties with Increase of the Voters in Indonesia General Election 2014

| Rank | Political Party | Elections 2009 | Elections 2014 | Change |
|---|---|---|---|---|
| 1 | PDIP | 14.03% | 18.95 % | + (↑) |
| 2 | PGolkar | 14.45% | 14.75 % | + (↑) |

| Rank | Political Party | Elections 2009 | Elections 2014 | Change |
|------|----------------|----------------|----------------|--------|
| 3 | PGerindra | 4.46% | 11.81 % | + (↑) |
| 5 | PKB | 4.94% | 9.04 % | + (↑) |
| 6 | PAN | 6.01% | 7.59 % | + (↑) |
| 8 | PNasdem | - | 6.72 % | - |
| 9 | PPP | 5.32% | 6.53 % | + (↑) |
| 10 | PHanura | 3.77 % | 5.26 % | + (↑) |
| 12 | PKPI | 0.90 % | 0.91 % | + (↑) |

Among twelve political parties, eight of them succeed to increase to percentage compared to 2009 legislative election (Table 4), one new political party (Table 4), but three of the political parties has decreased the number of electors (Table 5).

Table 5. The List of Political Parties with Decrease of the Voters in Indonesia General Election 2014

| Rank | Political Party | Elections 2009 | Elections 2014 | Change |
|------|----------------|----------------|----------------|--------|
| 4 | PDemokrat | 20.85% | 10.19 % | - (↓) |
| 7 | PKS | 7.88% | 6.79 % | - (↓) |
| 11 | PBB | 1.79 % | 1.46 % | - (↓) |

Among those results, president candidates number 1 suppoted by seven political parties. Four of the seven political parties grouped as a political party revenue increased voice in the legislative elections of 2014. Whereas the other three political parties voice revenue decline. Otherwise, all of political parties that supporting president candidates number 2 are grouped as the political parties with increased voice.

## 3.2 President Candidates's Facebook Page

The first social media analyzed in this article is Facebook. The Facebook connectivity help the group to build a political party to further back up their main figure in the forecast Presidential candidacy [17]. Table 6 shows the popularity of every candidate in presidential election July 2014.

Table 6. Presidential candidates' Facebook Fans (July 2, 2014)

| No | Political Party | Like | Talking about this | Most Popular Age Group | City |
|----|----------------|------|--------------------|------------------------|------|
| 1 | Prabowo Subianto | 7,257,921 | 2.4M | 18-24 | Jakarta |
| 2 | M. Hatta Rajasa | 120,783 | 43.1K | 18-24 | Jakarta |
| 3 | Joko Widodo | 2,998,487 | 1.6M | 18-24 | Jakarta |
| 4 | M. Jusuf Kalla | 1,037,797 | 210.9K | 18-24 | Jakarta |

Author found that Prabowo Subianto, president candidate number 1, is the most popular candidate in Facebook. This research also confirm that dominant Facebook's users that like political parties are young adults (19-24), and the most populous Facebook's city is Jakarta.

## 3.3 President Candidates's Twitter Account

One of number one activity on the web is Twitter [10]. Twitter is a microblogging website where users read and write millions of short messages on a variety of topics every day[9]. The analysis suggests that politicians are attempting to use Twitter for political engagement, though some are more successful in this than others[18].

Author capture the numbers of tweets, photos/videos, followers, and favorites of every president candidate (Table 7).Author found that Joko Widodo is president candidate with the most followers in Twitter, followed by Prabowo Subianto, Jusuf Kalla, and Hatta Rajasa.

Table 7. Presidential candidates' Twitter's Tweets, Photos/Videos, Followers, and Favorites (July 2, 2014)

| No | Political Party | Tweets | Photos/ Videos | Followers | Favorites |
|----|----------------|--------|----------------|-----------|-----------|
| 1 | Prabowo Subianto | 8,165 | 347 | 950,658 | 21 |
| 2 | Hatta Rajasa | 1,179 | 14 | 636,968 | 14 |
| 3 | Joko Widodo | 927 | 24 | 1,676,380 | 12 |
| 4 | Jusuf Kalla | 2,710 | 634 | 929,536 | 15 |

## 3.4 Social Media as President Candidates's Virtual Society

Social media like facebook and twitter create a new virtual world with the same interest. In presidential campaigns, social media has the ability to engage huge number of online users to interact each other in fostering the growth of virtual community and public spere [19]. On the other hand, [9] the users like twitter's followers will share the same political information with their connection. They create the virtual society via links of the themes of any tweets. And new online media like facebook or twitter become the channel for this purposes [20].

Author uses online social media research domain (SocialBaker) to find out the top online community in Indonesia. The supporters of the political party presidential candidates have different ways of expressing support for the presidential candidate. Table 8 lists the rank of every president candidate fans.

| # | | Page | Local Fans ▼ | Fans |
|---|---|------|-----------|------|
| 1 |  | Prabowo Subianto | 6 713 631 | 7 227 798 |
| 2 |  | Partai Gerakan Indonesia Raya… | 2 924 383 | 3 094 749 |
| 3 |  | Susilo Bambang Yudhoyono | 2 800 193 | 3 012 922 |
| 4 |  | Joko Widodo | 2 742 236 | 2 954 500 |
| 5 |  | Barack Obama | 1 504 520 | 41 366 988 |
| 6 |  | Gita Irawan Wirjawan | 1 381 901 | 1 408 583 |
| 7 |  | Informasi Beasiswa S1 S2 S3… | 1 183 451 | 1 227 009 |
| 8 |  | MUH. JUSUF KALLA | 938 735 | 1 019 610 |
| 9 |  | Ibu & Aku | 865 514 | 888 052 |
| 10 |  | B.J. Habibie | 828 490 | 863 516 |

*Figure 4. Indonesian Society Facebook Page (May 2014)*

The rank of presidential candidate popularity is lead by the president candidate number 1, Prabowo Subianto, followed by president candidate number 2, Joko Widodo. But for vice president candidate, the popularity of Jusuf Kalla, vice president candidate of Joko Widodo, reach better fans compared to Hatta Rajasa.

Table 8. President and Vice President Candidates' Society Fans (SosialBakers) Before the Election Day

| Society Rank | Icon | Page | Status | Local Fans | Fans | International Fans |
|---|---|---|---|---|---|---|
| 1 |  | PrabowoSubianto | President candidate | 6713631 | 7227798 | 514167 |
| 4 |  | JokoWidodo | President candidate | 2742236 | 2954500 | 212264 |
| 8 |  | MUH. JUSUF KALLA | Vice President candidate | 938735 | 1019610 | 80875 |
| 66 |  | Muhammad HattaRajasa | Vice President candidate | 100967 | 119672 | 18705 |

### 3.5 Real Count Results for Presidential Election 2014

After the vote count takes place from the 9[th] of July 2014, the Commission (KPU) finally officially announced the results of the real count. Commission (KPU) decided that the couple Joko Widodo and Jusuf Kalla is the winner of the 2014 presidential election. Jokowi represent the voters of the most populous island, Java. Meanwhile, Jusuf Kalla representing the voters of eastern Indonesia.

The current result already predicted by many people not only in Indonesia but also in International community. It seemed the power of social media has ability to promote online campaigns for presidential election 2014. It is also predict that in the future is expected to be a major media campaign.

## 4. CONCLUSIONS

Social media as one of internet application offers many benefits to various of social life. It has changed the way of political parties campaigns' styles and approaches. According to information from sections above, Author highlighted some important points of presidential candidates' campaigns on social media in Indonesian presidential election 2014

1) All of Indonesian presidental candidates already aware of the power of social media for their political campaigns.
2) Facebook is still the best social media for every presidential candidates' cyber campaigns.
3) Prabowo, presidential candidate number 1, is the most populer president candidate in Facebook not only in Indonesia but also from International fans.
4) Joko Widodo, president candidate number 2, is the most followed president candidate in Twitter. He become the winner based on KPU announcement (July [22]nd, 2014)
5) Dominant online user in Indonesian' facebook are young adult (18-24).
6) Jakarta is the city that consist of many
7) In fact to the real count announcement by KPU (July 22[nd], 2014), Joko Widodo and Jusuf Kalla won the presidential competition for 2014.
8) For next research, author interested to investigate the role of social media for local campaigns/elections, ASEAN community, etc.